\documentclass{scrartcl}

\usepackage{amsmath,amssymb}
\usepackage[noblocks]{authblk}
\usepackage{hyperref}
\usepackage{multicol}
\usepackage{multirow}
\usepackage[normalem]{ulem}
\usepackage{longtable}
\usepackage{booktabs}
\usepackage{array} 
\usepackage{calc}
\usepackage{graphicx} 

\usepackage[
style=authoryear,
natbib=true
]{biblatex} 
\usepackage{tabularx} 

\newcolumntype{M}[1]{>{\centering\arraybackslash}m{#1}} 

\usepackage[left=2cm, right=2cm, top=3cm, bottom=3cm]{geometry}

\title{Fate of nitrogen in French human excreta: current waste and agronomic opportunities for the future}

\author[1]{Thomas Starck\footnote{thomas.starck@polytechnique.org}}
\author[1]{Tanguy Fardet}
\author[1,2]{Fabien Esculier}
\affil[1]{LEESU, Ecole des Ponts, Univ Paris Est Creteil, Marne-la-Vallée, France}
\affil[2]{METIS, Sorbonne Université, CNRS, EPHE, Paris, France}

\date{}

\begin{document}

\maketitle

\section*{Abstract}

\begin{abstract}
Nitrogen (N) is essential for plant growth and protein synthesis but global reactive N losses, mainly from food systems, induce strong environmental impacts.
N losses after human excretion are often overlooked because, in Western societies, they partly occur as inert N\textsubscript{2}, following denitrification in wastewater treatment plants (WWTP), and losses in waters are often small compared to diffuse agricultural emissions.
Yet N from human excretions could be used for crop fertilization, potentially with very high recycling rates via source separation.
In this study we use unique operational data from the $\sim$20,000 French WWTPs to produce a N mass-balance of excretions in the French sanitation system.
Even though 75\% of WWTPs' sludge is spread on crops, only 10\% of the excreted N is recycled and 50\% of N is lost to the atmosphere, mainly through WWTP nitrification-denitrification.
The remaining 40\% ends up in water or in diffuse losses in the ground, of which about half is lost outside of the WWTPs' discharge system, through sewers storm water and individual autonomous systems.
While WWTPs removal efficiency increased in the 2000s, it has been followed by a decade of stagnation, reaching 70\% at the national level.
This national average hides regional discrepancies, from 60 to 85\% in the 6 French water agencies basins.
These differences closely correlate with the classification as ``N sensitive areas'' and is mainly due to large WWTPs which handle most of the N load.
Recycling all N in excretions could supply 10\% of domestic protein consumption in the current French food system, and up to 30\% if it is prioritized towards crop production for human consumption.
Redesigning the food system (decrease of nutrient losses, more plant-based diets) could further increase this contribution.
\end{abstract}

\pagebreak

\section*{Graphical abstract}

\begin{center}
    \includegraphics[width=0.8\linewidth, trim = {0 0 0 0}, clip]{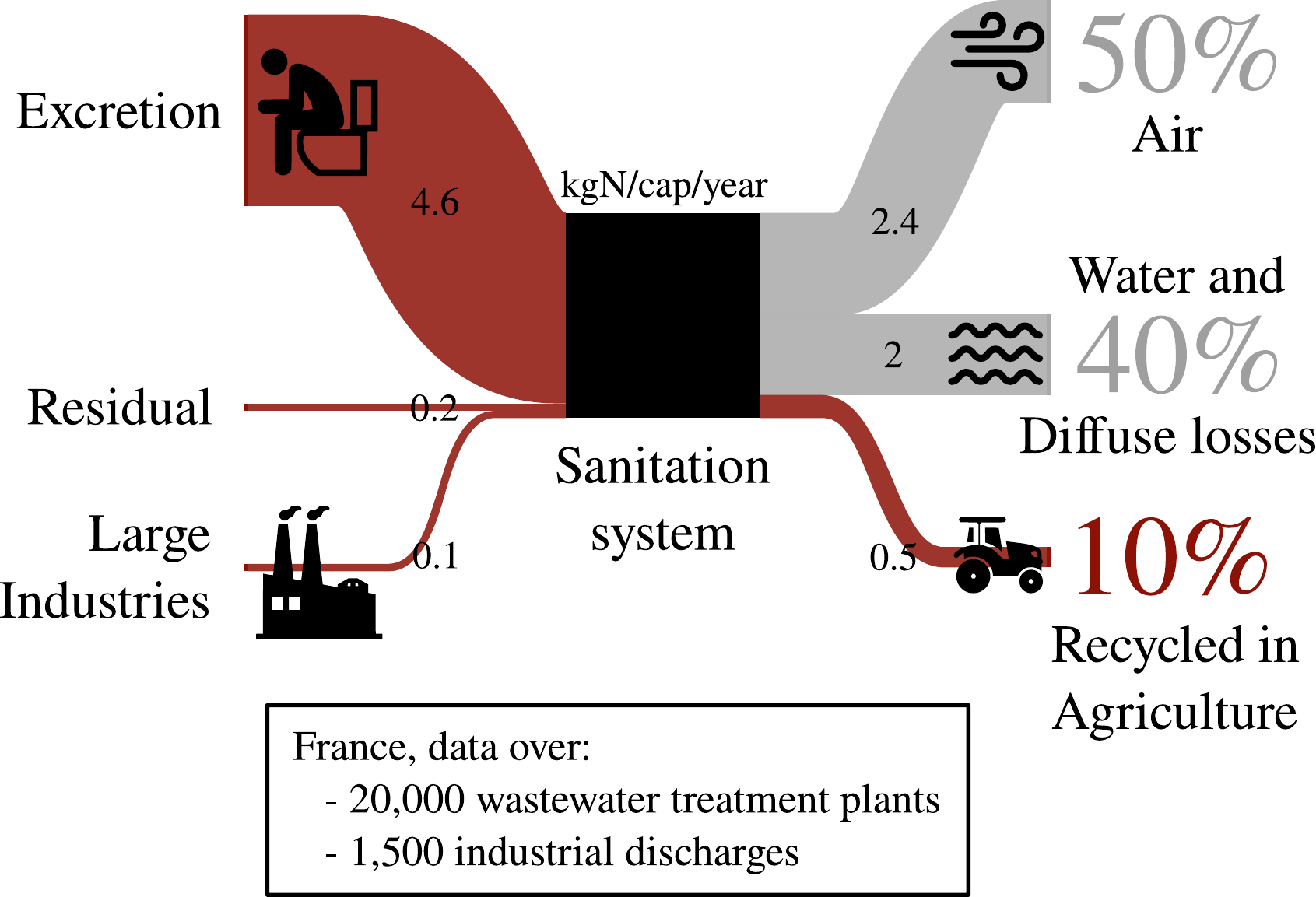}
\end{center}

\vspace{1em}

\section*{Highlights}
\begin{itemize}
\item We focus on the fate of N after human excretions in French wastewater.
\item Nitrogen (N) is barely recycled in wastewater treatment plants (WWTP).
\item We use multiple datasets to produce an N-budget of the French sanitation system.
\item $\sim$$90\%$ of excreted N is lost ($\sim$$50\%$  in air, $\sim$$40\%$ in water), and $\sim$$10\%$ is spread on crops.
\item Recycling all N in excretions would cover $10-30\%$  of French protein consumption.
\end{itemize}

\section*{Keywords}

Nutrient, Wastewater Treatment Plant, Sludge, Excretions, Denitrification, Food system

\section*{Abbreviations}

WWTP: Wastewater Treatment Plant; N: Nitrogen; NUE: Nitrogen Use Efficiency

\pagebreak

\tableofcontents

\pagebreak

\section{Introduction}\label{sec:introduction}

Compared to natural biological nitrogen (N) fixation, human activities and the discovery of the Haber-Bosch process doubled the global N input into the terrestrial biogeochemical flows.
This improved previously limited crop yields and enhanced soil carbon storage in some regions, but cascading N losses \citep{Galloway2003nitrogen} led to air and soil pollution, eutrophication, global warming, and biodiversity loss \citep{Fowler2013global, Gruber2008Earth-system, Vitousek1997Human}, also impacting human health \citep{Townsend2003Human}.
Nowadays, more than half of the proteins eaten by humans are obtained from nitrogen generated by the Haber-Bosch process \citep{Erisman2008How, Smil2001Enriching}.
This process experienced a 9-fold increase since the 1960s and now represents 1 to 2\% of global energy consumption \citep{Sutton2013Our}.
These increased N flows have exceeded the capacity of the environment to absorb them, leading to what \citet{Steffen2015Planetary} qualified as an overshoot of the planetary boundary on biogeochemical flows.
The global doubling of N flows hides larger regional impact.
In the US, for instance, the rate is 5 times above biological fixation \citep{EPA2011Reactive}.

To bring N cycle back into the planetary boundaries, three main routes can be followed to reduce N losses: a) decreasing N inputs b) improving N use efficiency (NUE), as current application practices lead to significant nitrogen loss, and c) increasing nitrogen circularity through the recovery of nutrients so they can be brought back to agricultural parcels.
While livestock excretions are nearly systematically reused in agroecosystems, such an approach is barely undertaken for human excretions.

In western countries such as France, most wastewater treatment plants (WWTPs) operate in 3 steps: a primary treatment removes suspended solids and particulate organic matter, a secondary treatment removes dissolved biodegradable matter and part of the nutrients, and a tertiary treatment further removes phosphorus and nitrogen to comply with discharge limit thresholds.
Most of the energy consumption is associated with  secondary treatment, due to the required aeration to promote aerobic nitrification \citep{Tchobanoglous2003Wastewater}.
Nitrifying/denitrifying 1 kg of N consumes the same order of magnitude of energy as Haber-Bosch fixation \citep{Sutton2013Our}.
While energy recovery from urban wastewater can be valuable to operate WWTP, the potential represents at the very best only 1\% of global anthropic energy consumption \citep{Rittmann2013energy}.
More interesting is to recover N for food self-sufficiency to decrease dependency to synthetic fertilizers, but current denitrification processes do not enhance circularity since N is lost in the air as N\textsubscript{2}.
This is further supported by the fact that only focusing on N discharge from WWTP will only marginally improve total nitrogen water contamination since at the national scale most of the N pollution in watersheds (\textgreater90\%) comes from diffuse agricultural sources and not from domestic discharge \citep{Billen2013nitrogen, Seitzinger2010Global}.
This does not undermine the fact that there can still be room for improvement locally, when WWTP removal efficiencies are low.

Given the significant costs associated to WWTPs and the size of the N deposit in human excreta (around 2 Mt of nitrogen per year in the EU, compared to a 10 Mt consumption from inorganic fertilizer), a precise assessment of the associated flows provides a valuable opportunity to understand how nitrogen management could be improved.
In particular, they can then be compared with alternative pathways for nitrogen management \citep{Martin2022Human}.

There has been rough estimations of sanitation systems N flows as part of national N budgets assessment, where the sanitation system is not the core aspect of the studies \citep{Häußermann2021National, Hayashi2021Nitrogen, Hutchings2014nitrogen, Papangelou2021Assessing}.
These estimations are rarely based on WWTP operational data and rather on mean removal efficiencies or concentration at the national scale.
More detailed assessments have been proposed at smaller regional scales, focusing on a few WWTPs \citep{Esculier2019biogeochemical}.

In this work, we provide a national-scale N mass-balance, based on operational data from more than 20,000 French WWTPs.
To our knowledge, this is the first national-scale sanitation system N mass balance based on operational data from so many facilities over more than a decade.
We quantify the nitrogen flows into the atmosphere, the water, and the soil systems to evaluate the French N-recovery rate from human excretions and how much of the current agricultural production it could sustain.

\section{Materials and methods}\label{materials-and-methods}

In this section we first describe the large datasets used for our N budget (Table \ref{table_data_sources}).
We also present the individual parameters used as inputs for our modeling (Table \ref{table_N_parameters}).
In both cases we further discuss the uncertainties in the Appendix \ref{appendix-supplementary-methods}.
From these, a N mass balance is performed for each of the 6 French water agencies basins which are then gathered to obtain the national balance.
We also describe our method to estimate the domestic protein consumption that could be covered by reusing all excretions as fertilizers (Table \ref{table_footprint_potential}).
The results are part of a larger project assessing nutrient flows in the French sanitation system; the code to generate, cleanup, and analyze the data is available at \url{https://codeberg.org/TStarck/N_P_France_sanitation_system}.
The original datasets sources before cleanup, the consolidated data used for the analysis, the graphs and more information are available at \url{https://zenodo.org/record/7990171}.

\subsection{Dataset sources}\label{dataset-sources}

There are over 20,000 WWTP in France, and about 1,500 industrial facilities report discharging N in sewers.
We use large datasets to estimate N flows discharged in sewers by industries, WWTPs inflows and outflows, and the different WWTP sludge destinations (Table \ref{table_data_sources}).
We corrected the data when obvious outliers were detected.

\begin{table}
    \caption{Data sources used for the N budget.}
    \label{table_data_sources}
    \centering
    \resizebox{\textwidth}{!}{%
    \begin{tabular}{m{0.15\linewidth}m{0.4\linewidth}m{0.5\linewidth}}
        \toprule
            \multicolumn{2}{c}{\textbf{parameter}} & \textbf{source} \\ 
        \midrule
            \multicolumn{2}{l}{N discharge to sewers from each individual industry} & GEREP database, nonpublic extension of the open access georisque French database, part of the European Pollutant Release and Transfer Register (E-PRTR). \\
            \multicolumn{2}{l}{Population by age and sex for each French city} & \citep{INSEE2022Population} \\
            \multicolumn{2}{l}{Sludge production and destination for each WWTP. } & \citep{PortailAssainissementCollectif2023} \\
        \hline
            \multirow{7}{=}{N flows in and out of each individual wastewater treatment plants, for each water agency basin)} & Artois Picardie basin \newline years 1992-2021 & Artois-Picardie agency \href{https://www.artois-picardie.eaufrance.fr/cartes-et-donnees/les-donnees-sur-l-eau-du-bassin-artois-picardie/}{website} \\
             & Rhin-Meuse basin, \newline years 1996-2021 & Rhin-Meuse agency \href{https://rhin-meuse.eaufrance.fr/telechargement?lang=fr}{website}  \\
             & Seine-Normandie basin years 2015, 2016, 2018, 2020 & Etat des lieux (status reports) data, only for 2015, 2016, 2018, 2020 communicated through mail. We also have the SIAAP data over 2007-2021, consisting of the 6 largest WWTP representing 50\% of the basin nutrient flows.  \\
             & Loire-Bretagne basin \newline years 2005-2021 & Communicated by email, publicly shareable  \\
             & Adour-Garonne basin \newline years 2000-2020 & Adour-Garonne agency website, \href{http://adour-garonne.eaufrance.fr/catalogue/581d5f70-558c-49e4-8d77-5bd4fe974b62}{link} for nutrient flows discharge and \href{http://adour-garonne.eaufrance.fr/catalogue/42f43670-099d-11de-97dd-001517506978}{link} for WWTP description  \\
             & Rhône-Méditerranée basin \newline years 2009-2020 & Rhône-Méditerranée agency \href{https://www.rhone-mediterranee.eaufrance.fr/telechargements/bibliotheque-de-telechargement-de-donnees-sur-leau}{website}  \\
        \bottomrule                  
    \end{tabular}
    }
\end{table}

\subsubsection{Data corrections}\label{data-corrections}

The datasets for sludge production, industry discharges and WWTPs inflows and outflows were screened for obvious outliers.
For each of the 6 water agency basins, we aggregate the N flows at the basin scale for each year and look for unexpected peaks on specific years.
This allowed
to spot obvious outliers.
When it was possible to identify that the peak was simply due to a misplaced comma for that particular year (factor 10, 100, or 1,000), we corrected the entry; otherwise, the outlier was not considered in the analysis and the entry was discarded from the final dataset.

\subsubsection{Nitrogen in and out of wastewater treatment plants}\label{nitrogen-in-and-out-of-wastewater-treatment-plants}

There are more than 20,000 WWTPs in France.
We gathered data about the annual mean N inflows and outflows for each WWTP, based on data provided by the 6 French water agencies.
The temporal availability of the data varies for each basin, but spans at least one decade (e.g., 1992-2020 for Artois-Picardie, 2009-2020 for Rhône-Méditerranée) except for the Seine-Normandie basin, where data was only available for the years 2015, 2018 and 2020.
For this basin, we retrieved additional data from the SIAAP (Syndicat Interdépartemental pour l'Assainissement de l'Agglomération Parisienne), the syndicate in charge of 6 of the largest WWTPs in the Paris region, which handle half of the basin's pollution.
The SIAAP data provides partial data for 2007-2020.
Uncertainties ($\sim$10\%) are discussed in Appendix \ref{a1.-uncertainty-on-nitrogen-flows-in-and-out-of-wastewater-treatment-plants.}.

\subsubsection{Sludge production and destination}\label{sludge-production-and-destination}

Annual sludge production by WWTPs and their destination (direct spreading, compost, incineration, landfill) is reported for each facility in the French sanitation portal database \citep{PortailAssainissementCollectif2023}, and at the national scale in Eurostat \citep{Eurostat2022Sewage}.
Total sludge production $p_t$ is consistent over the whole period, around 1-1.1 Mt/y.
However, the sum of the reported destination quantities \(r_{t} = \sum_{i}^{}d_{i}\) does not cover the whole production (only 0.9-0.95 Mt were reported in 2018-2021 across all destinations).
Therefore, we extrapolate the total amount of sludge sent to each destination i from the total production quantities, \(d_{i}\frac{p}{r_{t}}\) , considering that the fate of missing sludge follows the same distribution as the reported quantities.
To compute the N-budget, we average destination data over the 2018-2021 period because older data is barely reported and inconsistent.
Uncertainties ($\sim$10\%) are discussed in Appendix \ref{a2.-uncertainty-on-sludge-production-and-destination}.

\subsubsection{Large industries discharge in sewers}\label{large-industries-discharge-in-sewers}

The industries N annual discharge to sewers was estimated using the GEREP database, provided by the Direction Générale de la Prévention des Risques from the Ministry of Ecological Transition.
The database reports around 1,500 industrial facilities discharging nitrogen to sewers.
More information and a discussion about the uncertainties ($\sim$10\%) can be found in Appendix \ref{a3.-large-industries-discharge-in-sewers}.

\subsection{N-budget parameters}\label{n-budget-parameters}

In this section we describe the different coefficients used to produce our N mass balance, and their uncertainties (Table \ref{table_N_parameters}).

\begin{table}[h]
    \caption{Parameters used for the N mass balance.}
    \label{table_N_parameters}
    \resizebox{\textwidth}{!}{%
    \begin{tabular}{m{0.18\linewidth}m{0.28\linewidth}M{0.2\linewidth}M{0.2\linewidth}}
        \toprule
            \multicolumn{2}{c}{\textbf{parameter}} & \textbf{value} & \textbf{source} \\ 
        \midrule
            \multicolumn{2}{l}{N:P ratio in sludge at basin and national scale} & 2 & \citep{Fuchs2014Effets} \\
        \hline
            \multirow{7}{=}{Direct discharge and losses from sewers at the basin scale ($\%$ of pollution entering the wastewater treatment plants)} & Artois-Picardie basin & $20\%$ & \multirow{7}{=}{\centering Water agencies data } \\
             & Rhin-Meuse basin & $20\%$ & \\
             & Seine-Normandie basin & $10\%$ & \\
             & Loire-Bretagne basin & $15\%$ & \\
             & Adour-Garonne basin & $7\%$ & \\
             & Rhône-Méditerranée basin & $7\%$ & \\
             & France & $10\%$ & Combined basins \\
        \hline
            \multirow{7}{=}{Total population (million) and share not connected to sewers} & Artois-Picardie basin & 4.8M, $15\%$ & \multirow{7}{=}{\centering Water agencies data } \\
             & Rhin-Meuse basin & 4.3M, $6\%$ & \\
             & Seine-Normandie basin & 19M, $7\%$ & \\
             & Loire-Bretagne basin & 13M, $24\%$ & \\
             & Adour-Garonne basin & 7.8M, $30\%$ & \\
             & Rhône-Méditerranée basin & 16M, $25\%$ & \\
             & France & 64.9M, $18\%$ & Combined basins \\
        \hline
            \multirow{3}{=}{IAS N mass balance: relative destination} & air & $22\%$ & \multirow{3}{=}{\centering \cite{Catel2017Inventaires, Risch2021Applying}} \\
            & underground diffuse losses & $73\%$ & \\
            & sludge & $5\%$ & \\
        \hline
           \multicolumn{2}{l}{\multirow{2}{*}{\shortstack[l]{For people not connected to sewers, share of excretions in sewers \\ (i.e. excretion in public spaces connected to sewers).} }} & \multirow{2}{*}{1/3} & \multirow{2}{*}{estimation} \\ 
           \multicolumn{2}{l}{} & & \\   
        \hline
            \multirow{10}{=}{protein ingestion by French people (g/day)} & female, $\leq$10 years old & 56 & \multirow{10}{=}{\centering INCA3 study \citep{dubuisson2019third, data.gouv2021Données}} \\
             & female, 11-18 years old & 68 & \\
             & female, 18-44 years old & 71 & \\
             & female, 45-64 years old & 74 & \\
             & female, $\geq$65 years old & 69 & \\
             & male, $\leq$10 years old & 58 & \\
             & male, 11-18 years old & 84 & \\
             & male, 18-44 years old & 94 & \\
             & male, 45-64 years old & 96 & \\
             & male, $\geq$65 years old & 91 & \\
        \bottomrule                  
    \end{tabular}
    }
\end{table}

\subsubsection{Nitrogen excretion (urine and feces) from population}\label{nitrogen-excretion-urine-and-feces-from-population}

We use INSEE data describing the French population by city and by age \citep{INSEE2022Population}.
This is then coupled with data on N ingestion and excretion by age to determine N excretions for each water agency basin.
The population uncertainty is negligible compared to the other parameters.
However, there is a 3\% uncertainty at the national level as we do not account for tourism and demographic changes; this is discussed in Appendix \ref{a4.-population-and-tourism}.
Data related to protein ingestion by French citizens is reported in the INCA3 study \citep{dubuisson2019third, data.gouv2021Données}, detailed by age and sex categories.
We convert the protein contents to nitrogen with a 6.25 factor \citep{Maclean2003Food}.
We consider no N accumulation in adult bodies and consider that all ingested N is excreted \citep{Leach2012nitrogen}.
N excreted out of the sanitation systems has been neglected (diapers, skin and breath volatilization, open urination).

Combining both datasets gives the N excretion for each water agency basin and at the national scale.
This results in an averaged national N excretion of 4.6 kgN/cap/year.
More information and a discussion of the uncertainties can be found in section Appendix \ref{a5.-approximations-regarding-n-excretion}.

\subsubsection{Individual Autonomous Systems}\label{individual-autonomous-systems}

To compute the fraction of the population that is not connected to sewers and relies on individual autonomous system, we use the Eurostat figure of 18\% for metropolitan France \citep{Eurostat2023Population}.
For the basins of the 6 water agencies, we use the figures found in their different water agencies basins experts' assessments \textit{Etat des lieux}.
For people not connected to sewers, we consider that 1/3 of their excretion happens in public spaces connected to sewers (worplaces, schools, and other public spaces).

The N mass balance of individual autonomous system was estimated from a French report \citep{Catel2017Inventaires} with the following fate for N: 22\% in air, 73\% in underground diffuse losses and 5\% in sludge.
We assume that all the sludge is subsequently directed to WWTPs.
A discussion of the uncertainties can be found in Appendix \ref{a6.-fate-of-n-from-individual-autonomous-systems}.

\subsubsection{N in sludge (N:P ratio)}\label{n-in-sludge-np-ratio}

Since WWTP data only provides the nitrogen input (\(N_{i}\)) and the amount \(N_{w}\) of nitrogen rejected in the water, the flows of N to the atmosphere (\(N_{a}\)) and the sludge (\(N_{sludge}\)) are computed based on the phosphorus data: since there is no gaseous form for P, we know that \(P_{sludge} = P_{i} - P_{w}\).
We use an average N:P ratio of 2 in the sludge to determine \(N_{sludge} = {2P}_{sludge}\), then compute \(N_{a} = N_{i} - \left( N_{w} + N_{sludge} \right)\) (see Appendix \ref{a7.-uncertainty-for-wwtp-flows-of-nitrogen-to-sludge-and-air} for discussion and sources on the N:P ratio).
We consider a 30\% N volatilization during the composting process \citep{zhong2013emissions}.
For more details and a discussion of the associated uncertainty, see Appendix \ref{a7.-uncertainty-for-wwtp-flows-of-nitrogen-to-sludge-and-air}.

\subsubsection{Direct discharge losses}\label{direct-discharge-losses}

Part of the excreted nitrogen never reaches the WWTPs due to incorrectly connected pipes, sewer overflows, and sewer leaks.
Based on individual WWTP data and the different water agencies' \textit{Etat des lieux}, we obtain a national mean of 10\% for pre-WWTP loss, with relative uncertainty of 50\%.
More details are given in \ref{a8.-nitrogen-loss-before-wwtps}.

\subsubsection{Residual pollution}\label{residual-pollution}

From the reported N flows entering the WWTPs and our estimation of sewer losses we estimate the N flows entering sewers.
This quantity is higher than the sum of human excretions and large industries discharges in sewers, leaving a residual N pollution entering sewers.
This residual pollution can be due to several causes: domestic kitchen wastes in the sink, small industries not required to declare their discharge in sewers or runoffs entering combined sewers with stormwater.

The residual flow entering individual autonomous systems is assumed to be proportional to the excretion/residuals ratio of pollution entering sewers networks, considering that for people not linked to sewers one third of their excretions happen in places connected to sewers.

\subsection{Potential for food domestic consumption}\label{potential-for-food-domestic-consumption}

By comparing the amount of N in excretions to reported French N fertilizer consumption \citep{FAOSTAT2023Fertilizers}, we evaluate the extent to which global and French domestic protein supply could be supported if all excreted N were reused as fertilizers.
This analysis highlights the potential reduction in reliance on the Haber-Bosch process.

Given that France operates as an open food system \citep{Billen2018Two}, we employ the concept of N footprint as outlined by \citet{Esculier2019biogeochemical} for the Paris megacity in 2012.
This approach allows us to gain a deeper understanding of how human N excretions could potentially contribute to the food supply within France.

The calculation of the N footprint involves two main steps.
Firstly, we start with the N associated with the food supply (\(N_{supply}\)) of a given population.
Next, we determine the N footprint required to produce this food (\(N_{f}\)) by considering the nitrogen use efficiency (NUE): \(N_{f} = N_{supply}\)/NUE.
We also differentiate between plant-based and animal-based products since they exhibit distinct NUEs.
To estimate the maximum potential contribution of excreted N to food production, we also compute  the food self-sufficiency if N excretions are preferentially allocated to plant-based products for human consumption.
This approach is adopted because plant-based products have the highest NUE and, consequently, the lowest footprint.

We calculate this contribution at both the global level and for the Paris megacity (10 million inhabitants), which serves as a representative approximation for France.
Although there may be slight differences in the precise results at the national scale (e.g., N intakes of 4.6 kgN/cap/year in our study compared to 4.9 in \citet{Esculier2019biogeochemical}, we assume that they should be similar, given that the diet is expected to be comparable across the country.
All the values used as inputs for our analysis are derived from average values provided by \citet{Sutton2013Our} and \citet{Esculier2019biogeochemical} and are presented in Table \ref{table_footprint_potential}.

\begin{table}
    \caption{N footprint from plant and animal products in France and in the world, and associated potential contribution of N excretions to protein supply in the current food system.}
    \label{table_footprint_potential}

    \vspace{1ex}
    
    \centering
    \renewcommand{\arraystretch}{0.8}
    \small
    \resizebox{\textwidth}{!}{%
        \begin{tabular}{m{0.2\linewidth} M{0.4\linewidth} M{0.4\linewidth}}

               & \multicolumn{2}{c}{\textbf{\shortstack{Reported by the studies \\(input values)}}} \\ 
                & \shortstack{World (MtN)\\(mean years 2000-2010)\\ \cite{Sutton2013Our}} & \shortstack{Paris Megacity (ktN)\\(year 2012)\\ \cite{Esculier2019biogeochemical}} \\ 
            \hline
                 N in excretions / ingestions & 19 & 49 \\ 
            \hline
                \multirow{4}{=}{N in domestic food supply} & Vegetal 22 & Vegetal 20 \\
                 & Animal 6 & Animal 45 \\
                 & Fish 4 & Fish 7 \\
                 & Total 32 & Total 72\\
            \hline
                \multirow{3}{=}{N footprint (excluding fish)} & Vegetal 50 & Vegetal 27 \\
                 & Animal 100-230 & Animal 375 \\
                 & Vegetal + Animal 150-280 & Vegetal + Animal 402 \\
            \hline
                \multirow{3}{=}{N use efficiency (supply/footprint)} & Crop $43\%$ & Crop $75\%$\\
                 & Animal 3-6\% & Animal $12\%$\\
                 & Vegetal + Animal 10-19\% & Vegetal + Animal $16\%$\\
            \bottomrule
            
            &\multicolumn{2}{c}{} \\

            & \multicolumn{2}{c}{\textbf{\shortstack{Potential contribution of N excretions to domestic food supply \\(results computed from input values)}}} \\ 

                 & \shortstack{World (MtN)\\(years 2000-2010)} & \shortstack{Paris Megacity (ktN)\\(year 2012)} \\ 
            \hline
                \multirow{8}{=}{N in food (before urban waste) obtained with N excretions used as fertilizer} & \textit{Prioritizing vegetal production} & \textit{Prioritizing vegetal production} \\
                 & Vegetal 8.2 & Vegetal 20 \\
                 & Animal 0 & Animal 3 \\
                 & Total 8.2 & Total 23 \\
                 & & \\
                 & \textit{Not prioritizing vegetal production} & \textit{Not prioritizing vegetal production} \\
                 & Total 1.9-3.5 & Total 8 \\
            \hline
                \multirow{10}{=}{Share of domestic N consumption covered with excretions used as fertilizer} &\textit{Prioritizing vegetal production} & \textit{Prioritizing vegetal production} \\
                 & Vegetal $37\%$ & Vegetal $100\%$ \\
                 & Animal $0\%$ & Animal $6\%$ \\
                 & Total $26\%$ (incl. fish) & Total $31\%$ (incl. fish)\\
                 & & \\
                 & \textit{Not prioritizing vegetal production} & \textit{Not prioritizing vegetal production} \\
                 & Total 6-11\% (incl. fish) & Total $11\%$ (incl. fish)\\
            \bottomrule
        \end{tabular}
        }
\end{table}

\section{Results}\label{results}

\subsection{French sanitation system N budget and recycling}\label{french-sanitation-system-n-budget-and-recycling}

\begin{figure}[h!]
\centering
    \includegraphics[width=0.95\linewidth, trim = {0 0 0 0}, clip]{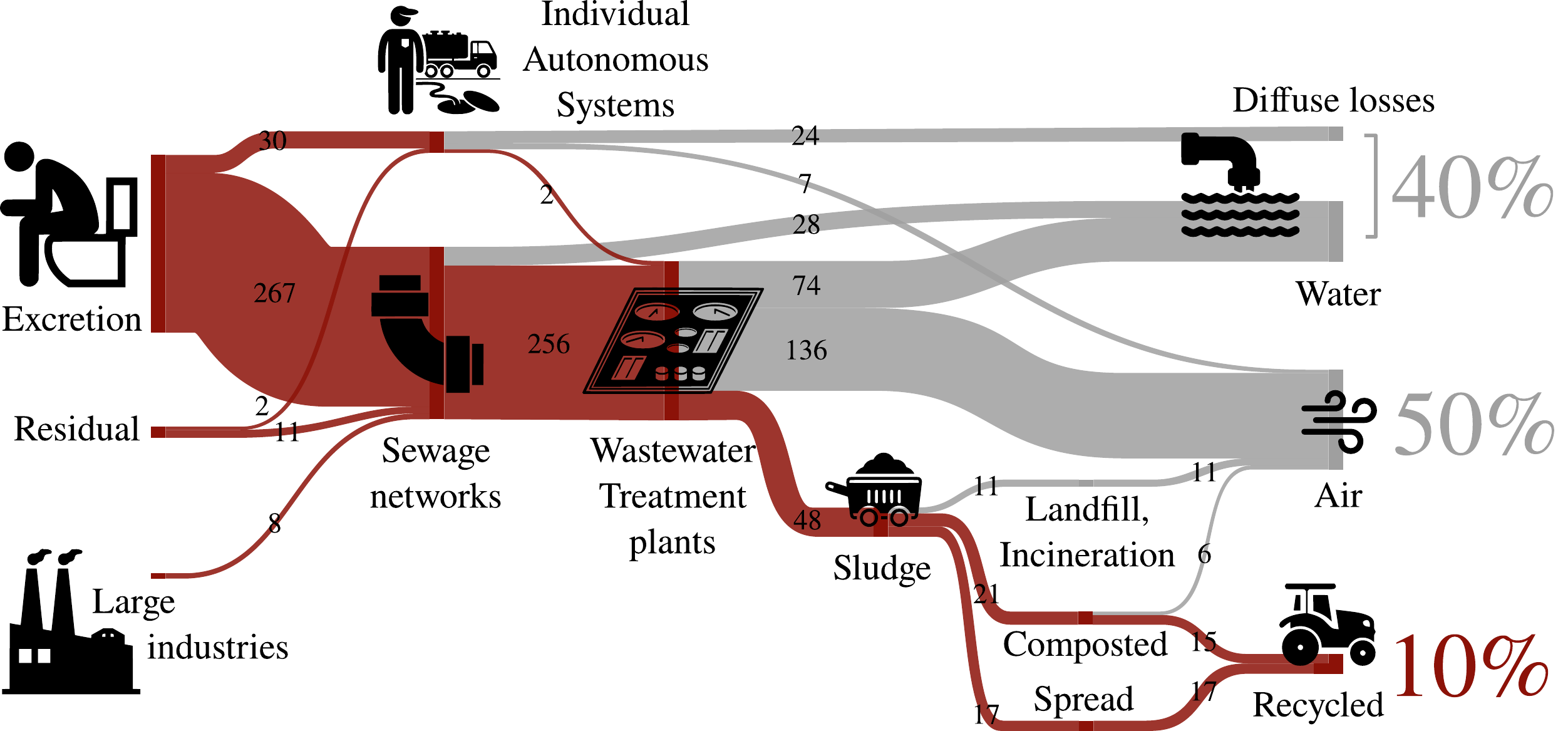}
    \caption{Yearly N flows in the French sanitation system (averaged for 2015-2020). Discrepancies between the sum of the incoming and outgoing flows are due to rounding. Unit: ktN.}
	\label{fig:sankey_N}
\end{figure}

For each water agency basin, we produce a N balance flow, using the computed quantities averaged over 2015-2020.
These are summed to produce the national N budget (Figure \ref{fig:sankey_N} for the precise numbers).

Over the course of a year, French N excretions are estimated to be about 300 ktN, large industries discharge in sewers represents $\sim$7.5\ ktN, and the residual N account for $\sim$12.5\ ktN (from kitchen waste, runoff entering combined sewers, small industries\ldots).
The 320 ktN from these 3 sources enter the sanitation system through sewers or individual autonomous systems.
In the end, 50\% (160 ktN) ends up in the air, mostly as N\textsubscript{2} following denitrification, 40\% reaches surface water or is diffusely lost in the ground (124 ktN).
Less than 10\% of the initial nitrogen (31 ktN) is recovered and used as fertilizer on crops.

85\% of the air emissions comes from nitrification/denitrification in WWTP, with the remaining 15\% coming from individual autonomous systems, sludge composting process, and sludge incineration.
Denitrification in WWTPs and losses outside WWTPs explain that, though 3/4 of French sludge are used to fertilize crops, the whole sanitation system recycling rate is only around 10\%.
Finally, despite the large relative uncertainty ($\sim$50\%) for the N contained in sludge used to fertilize crops, since the absolute recycling rate is low (\textless10\%), this translates in an absolute uncertainty of only a few percentage points, and the final recycling value is between 5\% and 15\%.

\subsection{WWTP removal efficiency through space and time}\label{wwtp-removal-efficiency-through-space-and-time}

\begin{figure}[h!]
\centering
    \includegraphics[width=0.95\linewidth, trim = {0 0 0 0}, clip]{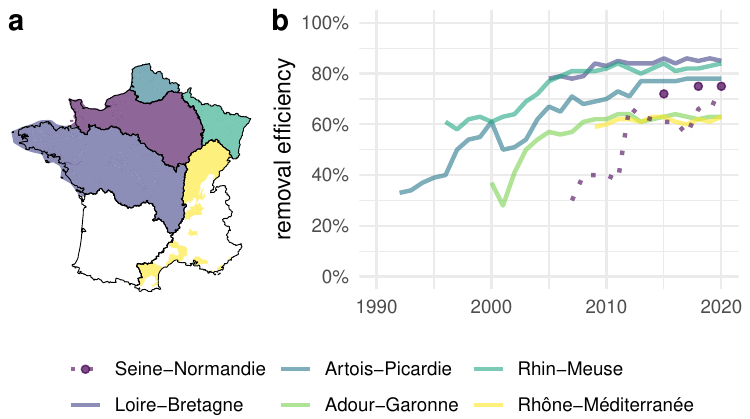}
    \caption{a) Nitrogen sensitive areas of the 6 French water agencies basins. b) Temporal evolution of N removal efficiency for the 6 basins. For Seine-Normandie, complete data was available only in 2015, 2018 and 2020 (points). The dotted line represents the partial data from the 6 largest Seine-Normandie facilities, handling about half of the total pollution.}
	\label{fig:N_basin_removal_efficiency}
\end{figure}

Removal efficiencies increased in all 6 French water agencies during the 2000s, and then leveled off in the 2010s, reaching a national mean of 70\%.
However, the final stagnation of that performance is closely correlated to the classification into N sensitive areas.
Adour-Garonne and Rhône-Méditerranée basins, largely considered ``non-sensitive to N'', have a global removal efficiency around 60-65\%, while the other basins reach 75-85\% (Figure \ref{fig:N_basin_removal_efficiency}).

\begin{figure}[h!]
\centering
    \includegraphics[width=0.95\linewidth, trim = {0 0 0 0}, clip]{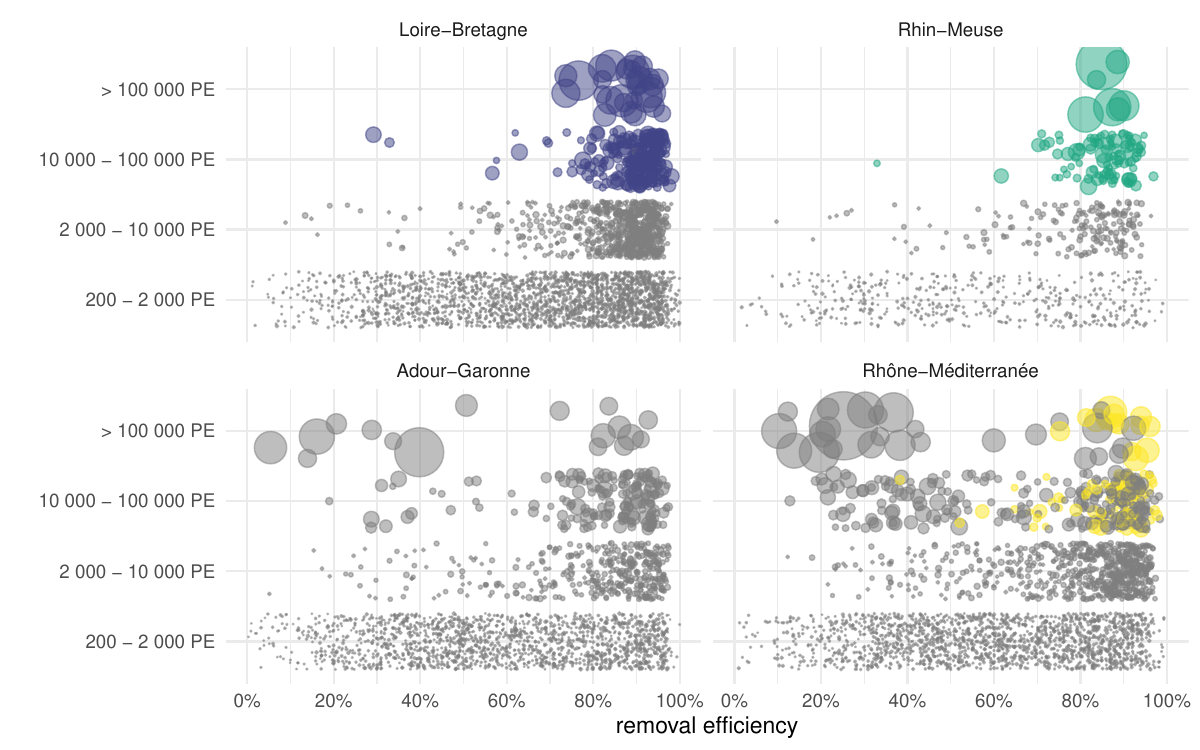}
    \caption{Individual WWTP N removal efficiencies, averaged over 2018-2020, in 4 of the French water agencies basins. Each dot is a WWTP, size is proportional to nominal capacity in population equivalent (PE). Dots are colored if the WWTP must reach minimal performances due to N sensitive area classification (see Figure \ref{fig:N_basin_removal_efficiency}a and text) or grey otherwise.}
	\label{fig:N_WWWTP_removal_efficiency}
\end{figure}

This difference is mostly due to removal efficiencies of large WWTPs.
Indeed, the EU regulation concerning WWTP in N sensitive zones requires a maximum outflow concentration of 10-15 mgN/l or an annual N removal efficiency of 70-80\% for facilities larger than 10,000 population equivalents (the French decree has translated the 70-80\% of the European directive \citep{EU2012Regulation} into a less protective 70\% \citep{Légifrance2015Arrêté}).
These facilities represent only 5\% of French WWTP but handle 80\% of the flows.
There is no automatic requirement regarding N for smaller WWTPs (\textless10,000 population equivalent), or the ones that are not in N sensitive zones. 
Figure \ref{fig:N_WWWTP_removal_efficiency} illustrates this effect of the European directive.
Facilities larger than 10,000 populations equivalent and in N sensitive areas generally have removal efficiencies above 70\%. 
On the contrary, large stations not classified as N-sensitive, e.g., in Adour-Garonne and Rhône-Méditerrannée, can have removal efficiencies well below 50\%. 
The map on Figure \ref{fig:N_basin_removal_efficiency}a also reveals that the criteria determining whether a basin is N sensitive have most probably other prevailing motivations than an ecological analysis of N sensitivity. 

Finally, WWTPs smaller than 2,000 population-equivalent take a wide range of values, but intermediate WWTPs (2,000 to 10,000) tend to have relatively high removal efficiencies even  though they are not subjected to legal requirements.

\subsection{Current agricultural potential}\label{current-agricultural-potential}

N fertilizers consumption in France is 2 MtN, and 120 MtN in the world.
With respectively 0.3 (this study) and 19 MtN contained in excretions \citep{Sutton2013Our}, human excretions could replace around 15\% of synthetic fertilizers consumption in both cases.

At the global scale and within the current food system, based on our N footprint method, we find that completely recycling human N excretions could cover 6-11\% of the protein demand\emph{.} Prioritizing the recycling to plant production for human consumption could supply up to 26\% of the protein demand.
In the case of France, the potential contribution to protein domestic supply is 11\% and up to 31\% if plant-based products are prioritized (cf. Table \ref{table_footprint_potential}).

\section{Discussion}\label{discussion}

\subsection{Consistency of the nitrogen flows}\label{consistency-of-the-nitrogen-flows}

Removal efficiency for N tertiary treatment reported in \citet{Degrémont2005Mémento} is 70-80\% (70\% at the national level in our study).
At the national scale, the estimations are 70\% in Germany \citep{Häußermann2021National}, 83\% in Belgium \citep{Papangelou2021Assessing} and 60\% for Japan \citep{Hayashi2021Nitrogen}.
\citet{Vigiak2020Domestic} propose N removal efficiencies of 25\% for primary WWTP treatment, 55\% for secondary and 80\% for tertiary, while \citet{Drecht2009Global} propose respectively 10\%, 35\%, and 80\%.
Given that 1-2\% of French urban wastewater is handled through primary, 15-20\% through secondary, and 80-85\% through tertiary treatment \citep{Eurostat2023Population}, this would give a national removal efficiency of 75\% for \citet{Vigiak2020Domestic} and 72\% for \citet{Drecht2009Global}, very close to the 70\% of our study.
Finally, \citet{Hutchings2014nitrogen} focused on Denmark between 1990 and 2010.
The authors report an increase over time in WWTP yield, similarly to our results for France.
Removal efficiency was estimated at 7\% in 1990, 50\% in 1995, 70\% in 2000 and almost 80\% in 2010.

Our N mass balance within WWTPs at the national scale is consistent with literature values.
For large European cities in the 2000s, assuming an 80\% removal efficiency through denitrification, \citet{svirejeva2011nitrogen} estimated an output repartition from the plants of 58\% to air, 21\% to water and 21\% to sludge, close to our figure of 53\%- 29\%-18\%.
The national WWTP N mass balance for Germany is 55\%-30\%-15\% \citep{Häußermann2021National}.
The results are slightly different for \citet{Hayashi2021Nitrogen}, who produced an N budget for Japan through 2000-2015.
In 2015, out of the 760 ktN input to WWTPs in the wastewater, 40\% was released in the air, 40\% in surface waters, and 20\% in sludge, given a national removal efficiency of 60\% in WWTPs.
In each of these examples, the share of N in wastewater ending in sludge is close to 20\%, and the repartition between water and air reflects the removal efficiency of denitrification.
There is a more important difference with \citet{Hutchings2014nitrogen} for Denmark.
They report that, in 2010, 70\% of N in wastewaters was denitrified, 21\% was discharged in waters, and only 8\% ended up in sludge.
Part of the difference can be due to higher removal efficiencies, but this principally highlights the high uncertainty (factor 2) concerning our N:P ratio to determine the N content of sludge (see Material and Methods, we use a N:P ratio of 2 whereas \citet{Papangelou2021Assessing} report a ratio of 1.25-1.5 for instance).

Finally, for the 2000s in Europe, \citet{Drecht2009Global} proposed a human N emissions in wastewater (domestic + industries) of 5.7 kgN/cap/year, higher than our value of 4.9 kgN/cap/year, but \citet{Hutchings2014nitrogen} for Denmark in 2010 report a value exactly identical to ours.
Most of this N loads ($>90\%$) are due to excretions, highlighting the influence of dietary nitrogen consumption on WWTP activity \citep{rautiainen2023decreasing}.
In our budget, the residual quantity is the N flow entering sewers that could not be explained by excretions nor by large industries discharge to sewers and is estimated to be 12.5 ktN.
This could be due to kitchen waste entering the sink, rainwater runoffs entering combined sewers, or small industries not declaring discharge to sewers.
If we consider that the residual quantity is mostly due to households, this would mean the N emissions not related to excretions would represent about 4\% of domestic emissions, in line with literature values consistently reporting rates below 10\% \citep{Friedler2013Wastewater, Vinnerås2002performance}.
However, given the uncertainty on the excretion input, most of these residual inputs may also simply come from an underestimation of the N contained in human excreta.

\subsection{Low recycling rates are the norm in industrialized countries.}\label{low-recycling-rates-are-the-norm-in-industrialized-countries.}

Other studies performed N budgets of the sanitation system, usually as part of the larger frame of national N budget.
They are generally less detailed, and N loads coming to WWTP or removal efficiencies are often part of the models assumptions, and not based on comprehensive operational data.
Losses from sewers and combined sewers overflows are almost never assessed.

For Belgium in 2014, \citet{Papangelou2021Assessing} used a model with an WWTP N removal efficiency of 80-85\% as input, and a connection to sewers rate of 81\%.
But contrary to France WWTP sludge are seldom used for agricultural purpose in Belgium: in Wallonie (1/3 of population), only 50\% is spread on crops and the practice is forbidden in Flandre.
This results in a final global N recycling rate of less than 3\% (only 1.3 ktN returned to crops compared to 40 ktN in municipal wastewaters).

The case of Germany was studied by \citet{Häußermann2021National} for the years 2010-2014.
Virtually every inhabitant (\textgreater95\%) is connected to sewers \citep{Eurostat2023Population} but only 25\% of produced sludge is reused in agriculture \citep{Eurostat2022Sewage}.
Since 85\% of the N in wastewater ends up in the air or in the hydrosphere (380 ktN incoming, 211 ktN denitrified and 114 ktN discharged), the best recycling rate would be 15\% but is further diminished below 5\% by the fact that only ¼ of sludge are reused in agriculture \citep{Eurostat2023Population}.

For Denmark, over two decades, N recycling rate to agriculture barely increased, despite national removal efficiency increasing from 20\% to 80\%.
The recycling rate was $\sim$0\% in 1990 and a mere 6\% in 2010 (2 ktN out of 32.7 entering WWTP) \citep{Hutchings2014nitrogen}.

At a more regional scale, \citet{Esculier2019biogeochemical} focused on Paris urban area (today 10 million inhabitants) from a historical perspective.
They found only marginal improvement on excreted N recycling, which has stayed around 5-10\% for the last decades, after a peak over 40\% around 1900.
Figures for France in 2006 reported in \citet{le2018biogeochemical} confirm the low  recycling rate we found.
They proposed an N budget for French agri-food system in 2006.
The rough estimation (based on N sludge content) was that 7\% (26 ktN out of 388 ktN) of excreted N was spread on crops, a figure similar to our 10\% for the situation more than a decade later.

These studies support that in a wide variety of industrialized countries where excretions are mostly handled with sewers and WWTP, the national recycling rate of excreted nitrogen is consistently below 10\%.

\subsection{Perspectives for the evolution of WWTPs}\label{perspectives-for-the-evolution-of-wwtps}

N removal efficiencies closely match classification into N sensitive areas and European directives requirements.
First, with the temporal increase in the 2000s for each basin and the stagnation levels in the 2010s (Figure \ref{fig:N_basin_removal_efficiency}); second with the effect on large WWTP when they are in sensitive areas.
This illustrates the point made by \citep{gorman2013story}:

``\emph{The ambient concentrations of some naturally occurring compounds (including nitrogen compounds), long regulated by natural processes alone, had now come to be regulated by human institutions.''}

In our N budget, N discharged to waters is 74 ktN (uncertainty 10\%) while combined losses from individual autonomous systems (24 ktN) and sewers (28 ktN) represent 52 ktN, with higher uncertainty (respectively 25\% and 50\%).
This means that at least 30\% and up to 50\% of the non-denitrifying N losses happen outside WWTPs.
Further improvements in WWTP removal efficiencies will only increase this relative share.
Thus, future regulation on N pollution may need to look beyond WWTPs and address the whole sanitation system.

Another option would be to shift the main goal of the sanitation from minimizing N discharge to maximizing N recycling to agroecosystems.
This will still lead to minimize N losses but will also enhance food security, reduce energy consumption and greenhouse gases emissions from WWTP denitrification and fertilizers production.

\subsection{Potential contribution to protein supply}\label{potential-contribution-to-protein-supply}

Since urine concentrates 85\% of N excretions, (Figure S7) and urine-based fertilizers have very similar properties to synthetic fertilizers \citep{Martin2021Physico-chemical}, the comparison of the N excretion deposit to fertilizers consumption (15\%) can indicate the decrease in dependance towards foreign providers of synthetic fertilizers and natural gas for the Haber-Bosch process.

Our method finds an N footprint of 250 MtN (50 MtN for vegetal and 200 MtN for animal production), coherent with \citet{Sutton2013Our} assertion that ``only 20-30\% of the N originally introduced in agricultural soils ends up in food for human consumption as cereals, vegetables and fruits, while the rest is used to sustain the livestock production''.

Our protein self-sufficiency figures concern the food system as currently designed.
They could be increased by reducing food losses and waste, increasing crop and animal NUE and shifting towards more vegetal diets.
While increasing crop NUE ($\sim$50\%) at the global scale is possible, this would be more challenging for France where it is already high (75\%).
The repartition between N animal and vegetal products supply are 1/3-2/3 at the global scale and 70\%/30\% in France, highlighting the impact of diet.
Finally, about 1/3 of the global food production is lost or wasted \citep{Gustavsson2011Global}, and reducing it would increase the full chain NUE and the contribution of N excretions to food supply.

\subsection{Urine diversion for fertilizer productions}\label{urine-diversion-for-fertilizer-productions}

In Western societies, low recycling rates of N excretions have not always been the norm.
In Paris, this rate exceeded 40\% in the beginning of the 20th century.
This rate was achieved by a combination of high recycling of human excreta collected from cesspools and sewage farms enabling to return to agricultural land the excreta of people connected to sewers \citep{Esculier2021Past}.
More recently, it is estimated that urban excretions recycling rate in China were as high as 90\% in 1980, 70\% in 1990, 40\% in 2000, and below 5\% after 2010.
For rural settings, the estimation was 95\% in 1980, 77\% in 2000 and just 50\% in 2010 \citep{Gu2015Integrated}.

Only focusing on end-of-pipe recycling would have some drawback.
First, domestic consumption of water is around 100-150L/cap/day, while urine and feces production is around 1-1.5L/cap/day, resulting in a dilution by a factor 100.
The dilution factor can even reach 1,000 in the case of combined sewers.
Recovering highly diluted nitrogen faces strong technological locks that hinder the sewer rationale for a circular economy of human excreta nitrogen (and other nutrients such as potassium or sulfur).
While sewers can be adapted to humid climate, a large part of the population lives in regions where high water consumption is not sustainable.
Finally, in all the scenarios developed by \citet{Drecht2009Global}, less than 50\% of the global population is connected to sewers in 2050.

\begin{figure}[h!]
\centering
    \includegraphics[width=0.95\linewidth, trim = {0 0 0 0}, clip]{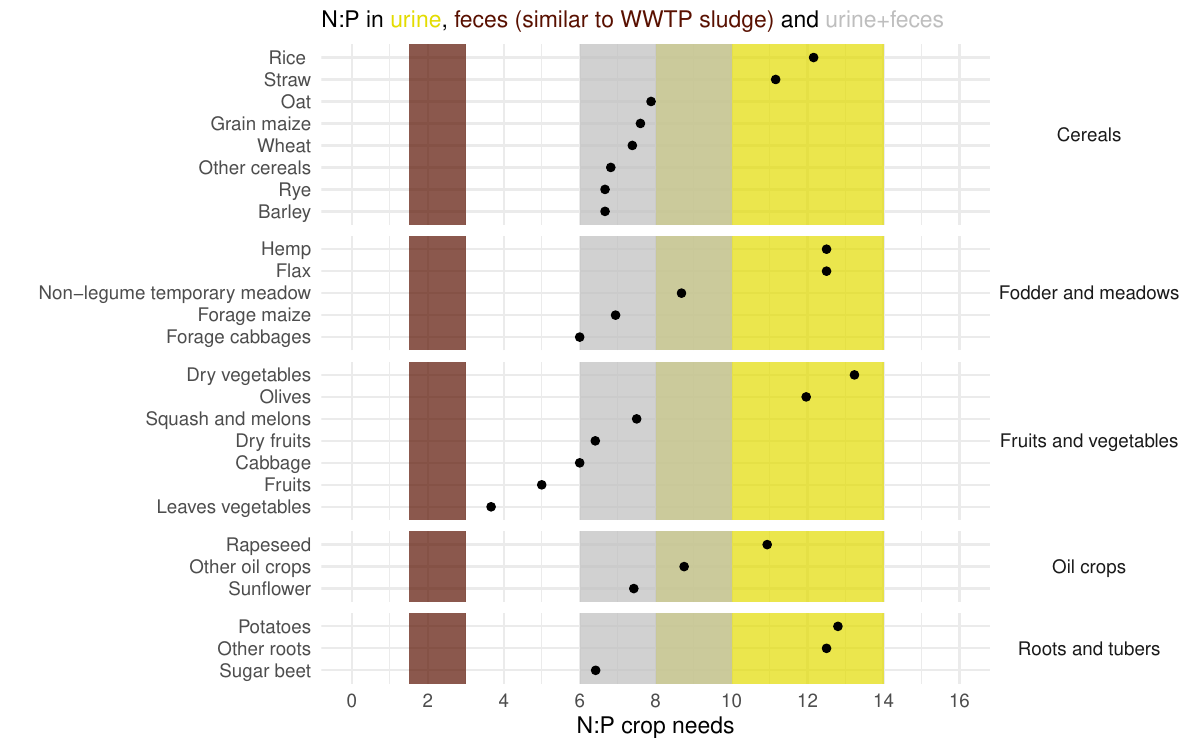}
    \caption{Comparison of crops N:P requirements (dots) to urine, feces and urine+feces N:P composition. Crops fertilization needs were determined using N:P composition reported in \citet{le2018biogeochemical} and a N use efficiency of 80\% \citep{Martin2021Physico-chemical}}
	\label{fig:ratio_crop_urine}
\end{figure}

The rationale for urine-based fertilizers have been described in seminal works as soon as the 1990s \citep{Larsen1996Separate, Larsen1997concept}.
Currently, WWTP nutrient removal produce sludge bringing human waste N:P ratio back to the value of feces, losing N through denitrification.
This N:P ratio of 1.5-3 is not adapted to crop requirements (Figure \ref{fig:ratio_crop_urine}).
Urine (containing 85\% of excreted N) diversion would yield fertilizers well-suited for crops needs, while greatly reducing the need for denitrification of wastewaters which would already have an N:P ratio close to the one targeted in sludge.
This would decrease WWTP energy consumption, mainly due to nitrification and denitrification.

Contrary to livestock breeding where the goal to minimize unproductive time leads to slaughtering just after the growth phase, most of human excretions happen during adulthood, at a time where virtually no nutrient is stored by the body.
As a result, human excretions are very similar to the ingested food and crops nutritional composition.
Livestock manure and slurry is excreted on the ground and mixed with bedding material, leading to high ammonia emissions in stables, often around 10-25\%.
On the contrary, when properly handled, losses during storage of human excretion are almost null \citep{Jönsson2013Closing}.

Finally, human urine has a low heavy metal contamination compared to sludge, but also to animal manure or biowaste \citep{Winker2009Fertiliser}.
One important challenge would be the storage, transport and application implementation of urine-based fertilizers, 2 orders of magnitude less concentrated than industrial fertilizers \citep{Larsen2013Implementation}.
An option is to use concentrated urine-based fertilizers.
Another important obstacle would be concerns concerning micropollutants \citep{winker2009pharmaceutical}.

\subsection{Limitations and perspectives}\label{limitations-and-perspectives}

The main limitations of our study are due to the uncertainty in the individual autonomous system N mass balance, the rough estimation of losses in sewers, and the N content of sludge.
We also could not appropriately estimate the N flow due to runoffs entering the combined sewers networks.
This issue is also present in other N-budgets and further investigation on this would be required to reduce the uncertainties around direct N losses upstream of WWTPs, especially they since may represent the main source of N pollution to water bodies from the sanitation system.

One of the most critical parameters for our French N recycling rate is the N quantity in sludge as opposed to the one denitrified.
Our comparison to literature review supports the general order of magnitude of our sludge / air repartition, but a large relative uncertainty remains (factor 2).
However, this translates in a small absolute uncertainty, and the French recycling rate is probably between 5 and 15\%.

\section{Conclusion and perspectives}\label{conclusion-and-perspectives}

We find that the vast majority (90\%) of N excreted by French people is lost, and only a small share (10\%) is finally reused in agriculture as fertilizers with WWTP sludge.
This happens despite WWTP  removal efficiencies of 70\% at the national scale.
Indeed, these facilities, while limiting N discharges to waters, can be seen as N destruction facilities.
Half of the N excreted by French people ends up in the air, mostly following the energy-intensive treatment in WWTP.
40\% of excreted N is lost in water or as diffuse losses.
Of these, 1/2 to 2/3 are due to WWTP discharges, and the rest is due to sewers and individual autonomous systems losses.

The N removal efficiencies of WWTPs increased in all basins before 2010, but have been stagnating ever since.
The performance achieved closely matches the regulatory classification as ``N sensitive areas'', with Southern basins (mostly not classified N sensitive) reaching $\sim$60\% removal efficiencies compared to 75-85\% for Northern basins (entirely classified N sensitive).
Tougher regulations can help decrease N water discharge but will not improve N recycling because of the inherently dissipative nature of WWTP N removal process.
Source separation of urine can help reach higher  rates, with fertilizers N:P composition more adapted to crops than current WWTP sludge.

The $\sim$300 ktN excreted by French people could currently cover about 10\% of the domestic protein consumption if they were recycled as fertilizers.
Prioritizing the recycling to plant-based production for human consumption could raise this share up to 1/3 of domestic consumption.
Changing the food system (more plant-based diets, lower overall protein intakes, increased N use efficiency in agrosystems) would further improve these figures.

\pagebreak

\section*{Funding}\label{funding}

Thomas Stack was awarded a PhD scholarship from Ecole Polytechnique and Ecole des Ponts ParisTech.
This project has received funding from the European Union’s Horizon Europe research and innovation programme under the Marie Skłodowska-Curie grant agreement No \href{https://cordis.europa.eu/project/id/101063239}{101063239} awarded to Tanguy Fardet.
The OCAPI Program is funded by several public institutions (\href{http://www.leesu.fr/ocapi}{www.leesu.fr/ocapi}).~

The funders had no role in study design, data collection and analysis, decision to publish, or preparation of the manuscript.

\section*{Credit authorship contribution
statement}\label{credit-authorship-contribution-statement}

\textbf{Thomas Starck:} Data curation, Formal analysis, Methodology, Funding acquisition, Writing - Original draft, Software, Visualization

\textbf{Tanguy Fardet:} Data curation, Validation, Funding acquisition, Writing - Original draft, Writing - review \& editing

\textbf{Fabien Esculier:} Supervision, Project administration, Methodology, Writing - review \& editing

\section*{Acknowledgements}\label{acknowledgements}

We thank the water agencies and the French Ministry of Ecological Transition for providing us with additional data that was not available in the French open-data repositories.
We also thank Gilles Billen, Josette Garnier, Julia Le Noë, Mathilde Besson, and Etienne Paul, for their advice and helpful discussions.
We thank the partners of OCAPI program for their support.

The Sankey diagramm (Figure \ref{fig:sankey_N}) was made using the Open-Sankey tool \url{https://open-sankey.fr/}.
On the Sankey diagram, the wastewater treatment plant icon is adapted from Daan\&apos's work; other icons were made by Luis Prado, Marco Livolsi, Jae Deasigner, Singlar, Gan Khoon Lay, WEBTECHOPS LLP, Trevor Dsouza, Elvn Sands and sandra from the Noun Project \href{https://thenounproject.com/}{https://thenounproject.com}.

\section*{Competing interests}\label{competing-interests}

The authors declare that they have no known competing financial interests or personal relationships that could have appeared to influence the work reported in this paper.

\pagebreak

\appendix

\section{Supplementary Methods}\label{appendix-supplementary-methods}

In the following we discuss with more details our methods for the N budget as well as data uncertainties.
When temporal series are available, we determine uncertainties by comparing the maximum absolute difference to the 2015-2020 mean (our reference period).
The figures illustrating these variations are available in the supplementary materials.

\subsection{Uncertainty on nitrogen flows in and out of wastewater treatment plants.}\label{a1.-uncertainty-on-nitrogen-flows-in-and-out-of-wastewater-treatment-plants.}

In the dataset used to quantify the N-flows, the largest stations, with a capacity \textgreater100,000 population equivalent, are monitored at least every week, and often every day \citep{Légifrance2015Arrêté}.
They  represent $\sim$50\% of the total French WWTP capacity.
Stations from 10,000 to 100,000 population equivalent, representing 30\% of the total capacity, are monitored at least every month.
So, for 80\% of the national flow, there is a high certainty about the yearly-averaged reported value.

After data cleaning, the year-to-year flows variability at the basin scale is small and is on the order of 10\% maximum (Figure S1 and S2).
We use this value as the uncertainty for incoming and outgoing flows.

\subsection{Uncertainty on sludge production and destination}\label{a2.-uncertainty-on-sludge-production-and-destination}

After correction, the year-to-year sludge production variability at the basin scale is small, on the order of 10\% maximum, except in 2016 for Seine-Normandie (Figure S3).
The relative destination of sludge is also rather constant starting 2018, with about 45\% of sludge composted and 30\% directly spread on crops at the national scale (Figure S4).
We use a 10\% uncertainty for sludge production and recycling.

\subsection{Large industries discharge in sewers}\label{a3.-large-industries-discharge-in-sewers}

We chose to use the non-open GEREP database rather than the open-access ``Géorisques'' database \citep{ministere2023} that also reports industrial facilities discharging pollutants as part of the European Pollutant Release and Transfer Register (E-PRTR).
``Géorisques'' data only reports facilities discharging more than 50 tons of N per year \citep{Légifrance2007Décret}.
On the other hand,the non-public GEREP data also includes facilities below this 50-ton threshold.
While ``Géorisques'' only reports a few dozen facilities discharging N in sewers, the GEREP database reports about 1,500 facilities, which increases the N flow by a factor 2.5. There is little year-to-year variability over the 2015-2020 period after data correction.
The maximum variability is around 20\%.

\subsection{Population and tourism}\label{a4.-population-and-tourism}

Between 2015 and 2020, the French population grew from 64.3 to 65.3 million residents.
In the analysis, we used the population census from 2018, with 64.8 million residents, hence a maximum error smaller than 1\%.

Every year, France receives around 70 million tourists who come to visit the country, staying on average a week\footnote{\url{https://www.insee.fr/fr/statistiques/1374552}}.
To this, one can add around 10 million tourists who transit through France from non-neighboring countries and around 100 million more who cross the border from a neighboring country, in both cases for a single day.
Taken together, this represents an increase of around 9 days worth of excretion per capita over a year.
This is partly compensated by the number of French residents going abroad for their holidays, which would represent around 250 millions days before COVID-19 so roughly 4 days per capita.

Overall it means that we may underestimate the national input to the sanitation systems from excreta by 2\%.

\subsection{Approximations regarding N excretion}\label{a5.-approximations-regarding-n-excretion}

The calculation of N excretion by the French population is subjected to two main sources of uncertainty: conversion from protein to nitrogen content and the approximation that all nitrogen intake is excreted.

For protein to nitrogen conversion, we chose to keep the ``standard'' 6.25 ratio as more recent publications criticizing this factor focused on the reverse operation (computing protein from nitrogen content), whereas the 6.25 ratio seems adequate to quantify nitrogen from animal protein \citep{Mariotti2008Converting}, which is the dominant protein source in France.
Searching for precise nitrogen to protein ratios in the literature we found only one reference quantifying nitrogen and amino acids in cereals \citep{Fujihara2008Nitrogen-to-Protein}, which gave ratios between 6 and 6.45 for the most common varieties eaten in France (rice, wheat, flour, oatmeal, quinoa), with the exception of corn flour at 7.

Regarding nitrogen retention in the body, we neglect it for adults, which should be a reasonable approximation on average.
We make the same approximation for children and adolescents, for which \citep{Jönsson2004Guidelines} found a mean storage rate of 2\% between 2 and 17 years.
This retention during growth translates into a mere 0.4\% total retention over a 80 year life.

For children below the age of 4, a large fraction of the excretions are collected into diapers and do not enter the sanitation system.
Overall, this should represent less than 2\% of the total excretions as this concerns the youngest children, i.e. those with the lowest nitrogen intake.

Part of the excreted N is also lost from other miscellaneous pathways (sweat, nails, hair, sloughed skin, body secretion), estimated at 8 mg/kg/day \citep{Pellet1990Protein}.
For a mean national body weight of 60-70 kg, this would give $\sim$0.5 gN.day\textsuperscript{-1}.cap\textsuperscript{-1} or 4\% of daily intakes, part of which would still go in sewers during shower.
On the one hand, overlooking miscellaneous N losses and N body storage  tend to overestimate N excretions by a few percent.
On the other hand, since the food intake survey might underestimate N intakes, we chose to neglect these losses.

In the end, our value for averaged N excretion at the national scale (4.6 kgN.year\textsuperscript{-1}.cap\textsuperscript{-1}, corresponding to 12.6 gN.day\textsuperscript{-1}.cap\textsuperscript{-1} or in terms of protein 78.5 g.day\textsuperscript{-1}.cap\textsuperscript{-1}) is in line with our literature review on the matter (Figure S1). For an N budget of Belgium in 2014, \citep{Papangelou2021Assessing} used an N consumption (and excretion) of 4.32 kgN.cap\textsuperscript{-1}.year\textsuperscript{-1}, close to our value of 4.6 kgN.cap\textsuperscript{-1}.year\textsuperscript{-1}

Considering the alignment of our values with the real excretion measures reported in the supplementary material, we estimate the uncertainty on nitrogen excretion to be $\sim$10\%.

\subsection{Fate of N from Individual Autonomous Systems}\label{a6.-fate-of-n-from-individual-autonomous-systems}

According to the Ministry of Ecological Transition, 15 to 20\% of the population is not connected to sewers and relies on Independent Wastewater Treatment \citep{Assainissement2023}.
Compared to the Eurostat figure of 18\%, which we use for metropolitan France \citep{Eurostat2023Population}, we estimate the uncertainty of this parameter to be around 15\%.

The N mass balance of individual autonomous system was estimated using LCA analyses of a septic tank with sand filter from \citep{Risch2021Applying} and a French report \citep{Catel2017Inventaires}.
The N destination for the \citep{Risch2021Applying}) is 3\% in air, 92\% in groundwater and 5\% in sludge but 22, 73, and 5\% respectively in \citep{Catel2017Inventaires}.
Since in both cases 5\% of the N ends up in sludge, the final global circularity figure is unaffected.
Only the repartition between groundwater and air emissions may change.
We estimate the uncertainty for the different N destinations to be 10\% for sludge, 25\% for groundwater and 85\% for air.

Part of nitrogen is also excreted as open urination directly on soil.
We did not specifically distinguish this portion, most probably very low, from excretions in individual autonomous systems.

\subsection{Uncertainty for WWTP flows of nitrogen to sludge and air}\label{a7.-uncertainty-for-wwtp-flows-of-nitrogen-to-sludge-and-air}

To determine the N composition of the sludge, we used a French collective scientific expertise review \citep{Fuchs2014Effets}.
When focusing only on studies reporting both N and P sludge content, the mean N:P ratio is 2.2. \citep{Papangelou2021Assessing} used a similar method but with an N:P ratio of 1.25-1.4 for their N budget in Belgium.
For our study we use a ratio of 2 and consider the uncertainty of our computed sludge N content to be 50\%.
This results in an absolute uncertainty of +/- 25 ktN for sludge N content (and deducted WWTP N\textsubscript{2} emissions) in our final N budget.
Since N\textsubscript{2} emissions from WWTP are 135 ktN in our final N budget, its relative uncertainty +/- 15\%.

Combining our N quantity with sludge production in tons of dry matter allows us to estimate mean N content, by basin and at the national scale.
We compare our values to the one reported in \citep{Fuchs2014Effets} (Figure S2).
We find similar results, even though our results are somehow a little higher than in the review.

\subsection{Nitrogen loss before WWTPs}\label{a8.-nitrogen-loss-before-wwtps}

Part of the N entering sewers does not reach the WWTP entrance, due to combined sewers overflows and sewers leaks.
Data are non-existent concerning this latter point, and very scarce for overflows.
Adour-Garonne and Loire-Bretagne basins report data from a sample of the largest WWTPs, where direct discharges represent respectively 7\% and 15\% of the flows finally entering the WWTPs.
For these 2 basins, we extrapolate the loss rate of this WWTPs sample to the whole basin.
No data was available for the Rhône-Méditerranée basin, so we used the loss rate of the other French Southern basin, Adour-Garonne.
For the three other basins (Rhin-Meuse, Seine-Normandie and Artois-Picardie), we use estimates from their respective experts' assessment \emph{Etat des lieux}.
Moreover, some nitrogen losses occur as ammonia volatilization from sewers but no estimations of theses losses at a sanitation system scale were found.

\printbibliography
\end{document}